\documentclass[10pt,conference]{IEEEtran}
\IEEEoverridecommandlockouts
\usepackage{multirow}
\usepackage{parcolumns}
\usepackage[T1]{fontenc}
\usepackage{stackengine}
\usepackage{pgfplots} 
\usepackage{tcolorbox}
\usepackage{listings}
\newlength\llength
\usepackage{comment}

\pgfplotsset{compat=newest}

\definecolor{mygr}{HTML}{e6e6e6}

\usepackage{listings}
\usepackage{xcolor}

\definecolor{trans}{HTML}{FDAE61}
\definecolor{rand}{HTML}{ABDDA4}
\definecolor{original}{HTML}{2B83BA}

\definecolor{airforceblue}{rgb}{0.36, 0.54, 0.66}
\definecolor{brown(web)}{rgb}{0.65, 0.16, 0.16}
\definecolor{backcolour}{rgb}{0.95,0.95,0.92}
\definecolor{copper}{rgb}{0.72, 0.45, 0.2}
\definecolor{cobalt}{rgb}{0.0, 0.28, 0.67}
\definecolor{arsenic}{rgb}{0.23, 0.27, 0.29}
\definecolor{burlywood}{rgb}{0.87, 0.72, 0.53}
\definecolor{bronze}{rgb}{0.8, 0.5, 0.2}
\definecolor{bondiblue}{rgb}{0.0, 0.58, 0.71}
\lstdefinestyle{code_list}{
    keywordstyle=\color{blue},
    numberstyle=\tiny,
    stringstyle=\color{brown(web)},
    basicstyle=\ttfamily\footnotesize,
    breakatwhitespace=false,         
    breaklines=true,                 
    captionpos=b,                    
    keepspaces=true,                 
    numbers=left,                    
    numbersep=5pt,                  
    showspaces=false,                
    showstringspaces=false,
    showtabs=false,                  
    tabsize=2,
    otherkeywords = { :,>=, <=, !=, True, False, ++},
    keywordstyle = [2]{\color{red}},
    morekeywords = [2]{length, equals, charAt},
    keywordstyle = [3]{\color{bronze}},
    morekeywords = [3]{checkPalindrome, check_palindrome, check_palindrome_helper},
    keywordstyle = [4]{\color{cobalt}},
    morekeywords = [4]{s, st, 1, 0, r, a, i,j},
    keywordstyle = [5]{\color{bondiblue}},
    morekeywords = [5]{String, True, False}
}

\lstdefinestyle{code_table}{
    backgroundcolor=\color{backcolour},   
    keywordstyle=\color{blue},
    numberstyle=\tiny,
    stringstyle=\color{brown(web)},
    basicstyle=\ttfamily\footnotesize,
    breakatwhitespace=false,         
    breaklines=true,                 
    captionpos=b,                    
    keepspaces=true,                 
    numbers=left,                    
    numbersep=5pt,                  
    showspaces=false,                
    showstringspaces=false,
    showtabs=false,                  
    tabsize=2,
    otherkeywords = { :,>=,++},
    keywordstyle = [2]{\color{red}},
    morekeywords = [2]{hasNextLine, nextLine, close, printStackTrace, println, System},
    keywordstyle = [3]{\color{bronze}},
    morekeywords = [3]{ readTextFile, read_textfile},
    keywordstyle = [4]{\color{cobalt}},
    morekeywords = [4]{sc, f, e, st},
    keywordstyle = [5]{\color{bondiblue}},
    morekeywords = [5]{String, Scanner, File, FileNotFoundException, with}
}

\usepackage{cite}
\usepackage{amsmath,amssymb,amsfonts}
\usepackage{algorithmic}
\usepackage{graphicx}
\usepackage{textcomp}

\newcommand{\jyoti}[1]{{\color{black} #1}}
\newcommand{\pinku}[1]{{\color{black} #1}}
\usepackage{hyperref}

\def\BibTeX{{\rm B\kern-.05em{\sc i\kern-.025em b}\kern-.08em
    T\kern-.1667em\lower.7ex\hbox{E}\kern-.125emX}}
\begin{document}

\title{Pathways to Leverage Transcompiler based Data Augmentation for Cross-Language Clone Detection\\

}

\author{\IEEEauthorblockN{ 
Subroto Nag Pinku} 
 \IEEEauthorblockA{\textit{Department of Computer Science} \\
 \textit{University of Saskatchewan}\\
  Saskatoon, Canada \\
 subroto.npi@usask.ca}
 \and
 \IEEEauthorblockN{Debajyoti Mondal}
 \IEEEauthorblockA{\textit{Department of Computer Science} \\
 \textit{University of Saskatchewan}\\
 Saskatoon, Canada \\
 d.mondal@usask.ca}
 \and
 \IEEEauthorblockN{Chanchal K. Roy}
 \IEEEauthorblockA{\textit{Department of Computer Science} \\
 \textit{University of Saskatchewan}\\
 Saskatoon, Canada \\
 chanchal.roy@usask.ca}

}

\maketitle

\begin{abstract}
Software clones are often introduced when developers reuse code fragments to implement similar functionalities in the same or different software systems resulting in duplicated fragments or code clones in those systems. Due to the adverse effect of clones on software maintenance, a great many tools and techniques and techniques have appeared in the literature to detect clones. Many high-performing clone detection tools today are based on deep learning techniques and are mostly used for detecting clones written in the same programming language, whereas clone detection tools for detecting cross-language clones are also emerging rapidly. The popularity of deep learning-based clone detection tools creates an opportunity to investigate how known strategies that boost the performances of deep learning models could be further leveraged to improve the clone detection tools. In this paper, we investigate such a strategy, data augmentation, which has not yet been explored for 
cross-language clone detection as opposed to single language clone detection. We show how the existing knowledge on transcompilers (source-to-source translators) can be used for data augmentation to boost the performance of cross-language clone detection models, as well as to adapt single-language clone detection models to create cross-language clone detection pipelines. To demonstrate the performance boost for cross-language clone detection through data augmentation, we exploit Transcoder, which is a pre-trained source-to-source translator. To show how to extend single-language models for cross-language clone detection, we extend a popular single-language model, Graph Matching Network (GMN), in a combination with the transcompilers and code parsers (srcML). We evaluated our models on popular benchmark datasets. Our experimental results showed improvements in F1 scores  (sometimes up to 3\%) for the cutting-edge cross-language clone detection models. Even when extending GMN for cross-language clone detection, the models built leveraging data augmentation 
outperformed the baseline with scores of 0.90, 0.92, and 0.91 for precision, recall, and F1 score, respectively.

\end{abstract}


\section{Introduction}
\jyoti{
Code clones are code fragments with similar functionalities in a software system. It appears when the developers reuse the existing source code knowledge base to annex the new features across the same or different platforms.  
Studies show that a software system may contain around 9\% to 17\% code clones\cite{zibran2011analyzing, baxter1997software}. Such clones have adverse impacts on software systems as they often introduce redundant codes, require code changes to implement consistently, and make program comprehension harder for the developers\cite{738528},\cite{1011328}. Moreover, reusing complex codes may generate intricate and sometimes incorrect programming logic. A rich body of research thus focuses on developing tools and techniques to detect and track clones across 
software systems so that they can be maintained through the complete software development life cycle\cite{hanjalic2013clonevol},\cite{mondal2019clone}. 
}

\jyoti{
In this paper we focus on clones that appear across software systems written in different programming languages. Clone detection techniques in such a cross-language context are relatively less explored compared to the decades of research on single-language clone detection\cite{tao2022c4}. Cross-language clones are ubiquitous in software systems that are written in one language with certain functionalities and need to be translated into another language to add support for multiple platforms. 
\pinku{
Clones carry important domain knowledge and thus studying the clones in a system could potentially assist in understanding the system itself \cite{johnson1994visualizing}. Cross-language clones  can play a crucial role in program comprehension due to their potential for revealing code reuse patterns across different languages. Analysis of code reuse patterns in cross-language context could help researchers understand development practices, identify toxic code snippets, and build code searching tools \cite{diamantopoulos2015employing} over diverse varieties of systems written in different languages. Furthermore, understanding the clones of a software system written in a certain programming language could potentially help understand a different software system written in a different programming language by tracking cross-language clones. Since multi-language software development (MLSD) appears to be common, at least in the open source world \cite{mayer2017multi}, cross-language clone detection could help understand any such clones and manage them accordingly in MLSD. Porting the same software systems to different platforms is natural and they could be written in different programming languages depending on the need. In such cases, cross-language clone detection tools will help detect, understand, and manage these clones across platforms. 
} 

Developers with existing knowledge and experience in one language make use of it to produce the same functionalities consciously or subconsciously. These codes are generally syntactically different but semantically the same\cite{mathew2020slacc}. Transforming systems written in one programming language to another programming language is a tedious task. These  can be considered as systems that are clones of each other whereas maintaining them over the years may take a lot of resources depending on their complexities. 
}

\jyoti{Detecting cross-language clone is difficult due to the difference in syntax and textual nature of source codes\cite{tao2022c4}. Figure \ref{fig:examplecross} shows an example of cross-language clones where three programs are written in two languages: the first one is in Java and the other two are in Python. All three programs solve the same problem of checking whether a given string is a palindrome. This example depicts the use of different structures and concepts such as library function, recursion, list slicing, etc. 
 Programs that are written in different languages differ inherently due to the grammar behind them. As a result, token-based, text-based traditional approaches tend not to work well for cross-language clone detection\cite{nichols2019structural}. Recently several deep learning models have shown good performances when detecting code clones in single-language settings\cite{zhang2019novel},\cite{wang2020detecting},\cite{ankali2021detection}. These models appear to have good capabilities of leveraging the underlying structure and semantics of the code fragments to identify code clones. Among many techniques proposed for single-language clone detection, some are based on graph neural networks that  
 consider both syntactic and semantic features of the code fragments. Consequently, such graph neural networks\cite{wang2020detecting} arguably learn the representation better than other deep learning methods that consider only syntactical features\cite{allamanis2018learning}.
}

\jyoti{A few techniques have been proposed for cross-language clone detection recently and deep learning models are shown to be effective in detecting clones\cite{nafi2019clcdsa, perez2019cross, tao2022c4}.  One of the challenges in building deep learning  models for cross-language clone detection techniques is the scarcity of data\cite{nafi2019clcdsa} as they typically   
require a large amount of data to well train the models\cite{yu2022data}. In general, the performance of deep learning models depends on the quality and amount of data used to train them. A common technique to boost the performance of deep learning models is to use data augmentation that adds new data to the  dataset by slightly modifying the existing data. Such data augmentation increases the number of data points and is shown  to enhance the generalizability of deep learning models\cite{shorten2019survey}. Although data augmentation techniques are well explored in computer vision and natural language processing research, only a few 
techniques are available that examine specifically the context of source code augmentation. These techniques predominantly focus on codes written in a single language and facilitate  rule-based data augmentation\cite{yu2022data}. 
While considering the problem of  cross-language clone detection, we noticed that a rich body of literature on transcompilers has gone unnoticed \cite{roziere2020unsupervised }. A natural question that we thought of is whether they could be leveraged to extend existing single-language clone detection tools for cross-language settings and moreover, for the purpose of data augmentation. Are there reasons to use transcompiler based augmentations instead of augmentations based on simple mutations (e.g., line deletion, swap, or duplication)? Considering the constraints of cross-language clone detection,   lack of sufficient data, and opportunities for leveraging transcompilers,  we formulated the following research questions for this study.\smallskip
}

\begin{figure}[t]
\begin{lstlisting}[xleftmargin=0.5cm,columns=flexible,language=Java,numbers=none, style=code_list]
public static boolean checkPalindrome(String st){
    String r = "";
    boolean a = false;
    int i = st.length() - 1;
    while(i >= 0){
        r = r + st.charAt(i);
        i = i - 1;
    }
    if(st.equals(r)){
        a = true;
    }
    return a;
}
\end{lstlisting}
\textbf{\footnotesize (a) Java code to check palindrome using library function} 
\begin{lstlisting}[xleftmargin=0.5cm,columns=flexible,language=Python, numbers=none, style=code_list]
def check_palindrome(s):
  def check_palindrome_helper(i, j):
    if j <= i:
      return True
    if s[i] != s[j]:
      return False
    return check_palindrome_helper(i + 1, j - 1)

  return check_palindrome_helper(0, len(s) - 1)
\end{lstlisting}
 \textbf{\footnotesize (b) Python code to check palindrome using recursion}\newline
\begin{lstlisting}[xleftmargin=0.5cm,columns=flexible,language=Python, numbers=none, style=code_list]
def check_palindrome(s):
    return s == s[::-1]
\end{lstlisting}
\textbf{\footnotesize (c) Python code to check palindrome using list  slicing}
\caption{Example of cross-language-clones: (a) is in Java,   (b)--(c) are in Python.
}
\label{fig:examplecross}
\end{figure}

\jyoti{
\noindent 
\textbf{RQ1.} \textit{\textbf{To what extent does source-to-source translation based data augmentation influence cross-language clone detection models?}}\\
In this research question, we explored the ability and extent of generalization for existing deep learning models through the lens of transcompiler-based data augmentation.\smallskip

\noindent
\textbf{RQ2.} \textit{\textbf{How does mutation based data augmentation perform compared to the source-to-source translation-based approach?}} \\
To answer this research question, we studied the effect of feeding a deep-learning model with codes generated through random modification. We compared models trained with this approach with the ones trained with the transcompiler-based augmented data.\smallskip

\noindent
\textbf{RQ3. }\textit{\textbf{Can we use source-to-source translation to adapt single-language clone detection models to detect cross-language clones?}}\\
To answer this research question, we investigated the opportunity to leverage existing single-language models for  cross-language clone detection in a combination with transcompilers. We chose the state-of-the-art model\cite{wang2020detecting} from the literature and extended it to cross-language settings.\smallskip
}

\jyoti{
\noindent
\textbf{Our contribution.} In this paper we answer the research questions RQ1--RQ3, which results into the following contributions:

 We introduce a data augmentation technique for cross-language clone detection using a transcompiler, which is a   pre-trained deep learning model for source-to-source translation. We conduct controlled experiments on the widely used CLCDSA dataset\cite{nafi2019clcdsa} with state-of-the-art deep learning 
    models for cross-language clone detection. Our experimental results show that the transcompiler-based data augmentation can boost the performances  of these clone detection models when trained with the augmented dataset (e.g., we noticed 3\% increase in F1 scores for some models).

    Since the transcompiler we exploit uses a semantic-preserving translation, we examined whether the same level of   performance boost could be achieved by   augmented datasets that are created using  simple mutation operations. By examining the abstract syntax trees  (ASTs) of the augmented data we show that the ASTs for transcompiler augmented data are more diverse which provides some insights and justification for its use in the cross-language clone detection setting.

    To examine whether transcompilers can aid in extending the single-language clone detection models for cross-laguage clone detection, we selected the Graph Matching Network (GMN), which is a widely used single-language clone detection technique and is known to show high performance on benchmark datasets\cite{wang2020detecting}. Given a pair of code fragments written in different languages, the idea here is to first use transcompilers to transform one of these code fragments to match the language of the other code fragment, and then to use a single-language clone detection tool on the new pair. However, the transformed code obtained from transcompilers may not always be parsable, and hence, it cannot be directly fed as an input to GMN. We tackle this challenge by 
    exploiting the srcML parser\cite{collard2013srcml}, which allows us to build XML representations to be used for GMN.  
    Our experimental results show that the performances of such  extended GMN models may not be the highest, but yet comparable to the cutting-edge cross-language clone detection models\cite{tao2022c4} that require high-end computing resources. This makes GMN an attractive option in a low-resource environment. Furthermore, this opens up the opportunity to explore whether the proposed framework for extending existing single-language clone detection models could be improved further or leveraged to build better cross-language clone detection models. 

}

\section{Background}\jyoti{
\subsection{Code Clone}
Codes that share syntactic and semantic similarities are clones of each other. 
Code fragments that are modified or transformed through editing have the same functionalities and are termed syntactic clones. 
Semantic clones are code fragments that have major differences in their structure and have the same meaning or semantics\cite{roy2007survey},\cite{al2020semanticclonebench}. 
Clones can be broadly divided into four types
\cite{roy2007survey},\cite{bellon2007comparison},\cite{svajlenko2015evaluating}. 
\textbf{Type-I}: Identical code fragments with a varying number of comments and white spaces.
      \textbf{Type-II }: Code fragments that are equivalent in syntax with changes in identifier names, literals,
types, layout, and comments.
     \textbf{Type-III}: Along with Type-I and Type-II, this type has addition, removal, and/or modification of statements.
       \textbf{Type-IV}: Code fragments that have the same functional behavior  but very different syntax. 

Software systems that are maintained across different platforms are often developed in different languages. Therefore, these systems often have code fragments written in different languages but with the same functionalities. Such code fragments are known as cross-language clones\cite{nafi2019clcdsa}. These clones can be categorized as Type-IV clones.

\subsection{Source-to-source translation}
Transforming codes from one programming language to another language is defined as source-to-source translation. It is often referred to as transcompilation
\cite{el2016enhanced}. 
There are a few open source packages such as java2python\footnote{https://github.com/natural/java2python}, chsarp2python\footnote{https://github.Scom/shannoncruey/csharp-to-python}, cs2j\footnote{https://github.com/twiglet/cs2j}, etc for source-to-source translation. These packages work only for the intended target language and are often unable to transform code the other way around. 
A few commercial tools are also available for high-level source-to-source 
translation\footnote{https://www.tangiblesoftwaresolutions.com/converters.html}. 
These tools and techniques rely on hand-crafted rules that use the mapping of keywords and libraries from one language to another. There exist  two well-known pre-trained deep learning models that can be used for code conversion. One is available through the   Microsoft  CodeXGLUE\footnote{https://github.com/microsoft/CodeXGLUE} project that can convert between C\# and Java code, and the other one is Transcoder, which is  available through Facebook research\footnote{https://github.com/facebookresearch/CodeGen} that can convert among Java, Python, and C++. 

\subsection{Pre-trained Models}
Pre-trained deep learning models have been successful in different natural language processing tasks. Among the available models, BERT\cite{devlin-etal-2019-bert}  and GPT\cite{brown2020language} achieved state-of-the-art results in different downstream tasks. 
With the inspiring results from these models, software researchers adopted them to build models such as CodeBERT\cite{feng2020codebert} and CodeGPT\cite{lu2021codexglue}. These models have been in use for different tasks such as code completion, code search, code summarization, and so on\cite{lu2021codexglue}.  


Transcoder is a pre-trained neural machine translation based model for code conversion among programming languages released by Facebook\cite{roziere2020unsupervised}. It uses the sequence-to-sequence modeling approach\cite{sutskever2014sequence} with unsupervised learning and exploits the transformer architecture with attention mechanism\cite{vaswani2017attention}. 
Transcoder achieved state-of-art accuracy in source-to-source code translation when compared with the existing commercial and non-commercial tools.

\subsection{Data Augmentation}
Deep learning models heavily rely on data to learn complex patterns from given data. With limited data, complex models are known to suffer from issues such as over-fitting and the inability to generalize to other datasets. Providing more data to the model is one of the fundamental ways to overcome such challenges\cite{shorten2019survey}. 
Creating synthetic data to increase dataset size by manipulating original data is known as data augmentation\cite{yu2022data}, which is widely used in computer vision and natural language processing domains. 

\subsection{Mutation Analysis in Code Clones}
Mutation is the process of modifying a piece of code to get another slightly modified version of it. Mutation analysis is primarily used in software testing\cite{bradbury2007comparative}. The primary goal is to introduce bugs in codes for testing. In code cloning, these operators are used to modify a code fragment to create another copy\cite{roy2009mutation}.  
Mutation analysis consists of a set of operations. These operations include renaming identifiers, insertion or deletion of a statement, inter-changing loops from one to another, swapping statements, etc. The mutation operations are language independent unless replacing one type of control with another, which is language-specific. Any of these operations can be used in any sequence to generate any number of copies of a code fragment. \pinku{Mutation analysis has been used to evaluate clone detection tools\cite{svajlenko2019mutation} and generate multiple copies of a code fragment\cite{fujiwara2019code}.}

\subsection{Graph Matching Network (GMN)}

Graph matching network is a graph neural network (GNN) that takes two homogeneous graph structures and finds the similarity between them\cite{li2019graph}. The goal of GNN model is to achieve embedding for the nodes of the graph by learning about the surrounding structure and semantics.  
Code fragments inherently have a structure that can be represented as trees or graphs. In representation learning, GNN learns from its surrounding neighbors and finds an embedding for each node. The GMN model has proved to be very effective for clone detection tasks in single-language settings\cite{wang2020detecting} with a proper embedding\cite{allamanis2018learning}. It takes the advantage of cross-graph attention to ensure that similar structure remains close in the embedding space and dissimilar structure spreads away while finding the global embedding. 
}
\section{Related Work}\jyoti{
\subsubsection{Single-Language Clone Detection}
Researchers have mostly been focused on single-language clone detection \cite{lei2022deep}. Traditional techniques primarily use structural features to detect single-language clones. Among them, a text-based approach such as  NiCad\cite{roy2008nicad} uses code normalization and text comparison to detect near-miss clones. Token-based approach SourcererCC\cite{sajnani2016sourcerercc} exploits tokens of code blocks to create an inverted index and compare them to find clones. However, it only captures information at the lexical level, which limits its performance. Deckerd\cite{jiang2007deckard} uses AST  to create clusters of numeral vector representations of subtrees. These subtrees are created from the AST of code fragments, and the cluster is created using the Euclidean distance metric. It is dependent on pre-defined language-specific rules. These techniques have been successful in detecting the first three types of clones. However, these conventional methods could not achieve the same performance for Type-IV clones\cite{roy2009mutation} due to the inability to capture the code semantics among code fragments with different syntax. 

The use of deep learning techniques has gained popularity because of its ability to learn the representation of the semantics of code fragments. White et al.\cite{white2016deep} applied deep learning to reduce the gap between the syntactic and semantics of code fragments by using both lexical information of identifiers and structural information of AST. DeepSim\cite{zhao2018deepsim} encodes Control Flow Graphs (CFG) to generate semantic metrics for deep neural networks. However, CFGs lack control and data flow information at different granular levels. Wang et al.\cite{wang2020detecting} addressed this issue and proposed a modified AST structure by adding additional control flow edges to the AST. The authors combined the modified AST with graph neural network 
to successfully detect clones. Another work by Phan et al.\cite{phan2017convolutional} used a graph-based convolution network with CFGs, which learns semantic features to find software defects. 

Ji et al.\cite{ji2021code} adapted a graph convolution network with hierarchical active graphs with an attention mechanism to distinguish the importance among nodes in the AST. Zhang et al.\cite{zhang2019novel} proposed an AST-based Neural Network (ASTNN), which splits ASTs into different sub-tree segments and uses bidirectional RNN. This method achieved success in both code classification and clone detection tasks. The recent success of encoder-decoder models in natural language processing tasks has paved the way for them in source code analysis applications. For instance, codeBERT\cite{feng2020codebert} and graphCodeBERT\cite{guo2020graphcodebert} are built on top of Google’s BERT model and have been used for multiple code tasks including clone detection.

\subsection{Cross-Language Clone Detection}
Only a few methods are known so far for cross-language clone detection due to the complexity and unavailability of proper datasets. Cheng et al.\cite{cheng2017clcminer} proposed CLCMiner, which uses code revision histories to detect clones. Their method did not use any intermediate representation and only relied on data from the version control system. LICCA\cite{vislavski2018licca}is another well-known tool that uses tree-based intermediate representation and variant of the longest common subsequence algorithm. Nafi et al.\cite{nafi2019clcdsa} used hand-crafted features to train a  siamese neural network for cross-language clone detection. The authors   provided a cross-language clone dataset in their paper\cite{nafi2019clcdsa}.

Mathew et al.\cite{mathew2020slacc} used dynamic analysis to detect cross-language clones. In another study, they studied clone detection as a special case of code search\cite{mathew2021cross}. This study combines a generic AST and token-based approach with non-dominating sorting. Tao et al.\cite{tao2022c4} proposed C4 that leverages contrastive learning and exploits the pre-trained codeBERT model. They experimented with the CLCDSA datasets and achieved state-of-the-art performance in cross-language clone detection.

\subsection{Source Code Augmentation}
With the advent of the deep learning era and the availability of many big and/or benchmark datasets, deep learning models have gained a lot of interest from the research community. Cross-language clone detection is one of the few areas that lack a proper dataset\cite{tao2022c4}. Existing techniques rely on small datasets and are often hand-crafted by the authors\cite{bui2018cross}. 

There exists a number of studies that applied data augmentation in the context of code clones. Yu et al.\cite{yu2022data} built a rule-based tool named SPAT for Java which has 18 transformation rules that can create semantically equivalent codes. The authors created these transformation rules by observing code patterns from clone pairs in the widely used BigCloneBench\cite{svajlenko2021bigclonebench} and OJ datasets\cite{mou2016convolutional}. The significant similarity in clone pairs enabled the authors to extract patterns from them.

Transforming codes to generate adversarial examples is often used to test a model's robustness. Zhang et al.\cite{zhang2020generating} used renaming techniques to attack code processing models. Models trained with these adversarial examples showed improvement in classification tasks and robustness of the model. Deepbugs\cite{pradel2018deepbugs} introduced a bug-inducing pattern in the code and proposed a name-based learning approach that detects bugs. It was very specific and limited to bug detections. Compton et al.\cite{compton2020embedding} trained a 
model with obfuscated codes. This obfuscation of variable renaming showed a decline in the performance of the model in the method-name prediction task though the embedding showed better preservation of semantics.

All existing research on source code augmentation used rule-based transformations that require manual investigation and human intervention. 
Such manual procedure often introduces systematic bias\cite{bird2009fair}, which jeopardizes the effectiveness of the process. Additionally, none of these transformation techniques focus on augmenting the dataset and are not built on top of any deep learning models. Our research is different from the studies mentioned in several aspects such as using the existing knowledge of transcompilers to augment data and extending single-language clone detection models for cross-language clone detection, which are both natural concepts that have not yet been explored in the literature. 
}

\section{Research Methodology} \jyoti{
\subsection{Problem Formulation}
We deal with two dimensions of cross-language clone detection: (a) the  clone detection task and (b) improving the deep learning based cross-language clone detection models through data augmentation.

Clone detection is formulated as a binary classification problem. Given   two code fragments $F_i$ and $F_j$, we define a clone pair $(F_i, F_j)$ and associate a label $L_{ij}$. A clone pair label is true when there are significant similarities between the fragments; otherwise, the label is false. A similarity score, $S_{ij}$ is calculated on the pair $(F_i, F_j)$, and the pair is true clone when $S_{ij}$ is larger than a 
threshold value\cite{svajlenko2014towards}. This similarity is based on the syntax or semantics of the code fragments. A pair of code fragments can only be called clones when we obtain a similarity score above the threshold. 

Improving a model's performance via data augmentation is formulated from the dataset perspective. 
Assume that $D$ is a given dataset $D$, and a model $M$ has an F1 score of $s$ on $D$. If we can augment the dataset $D$ by an amount $X$, train the model $M$ on $D+X$, and get an F1 score of $s+t$, then the difference in F1 score is  $t$, which we refer to as the improvement obtained by the augmentation. Note that in both cases the test dataset is the same, which is kept separate and thus  not augmented. 
}

 \begin{figure}[pt]
\centering
\includegraphics[width=.5\textwidth]{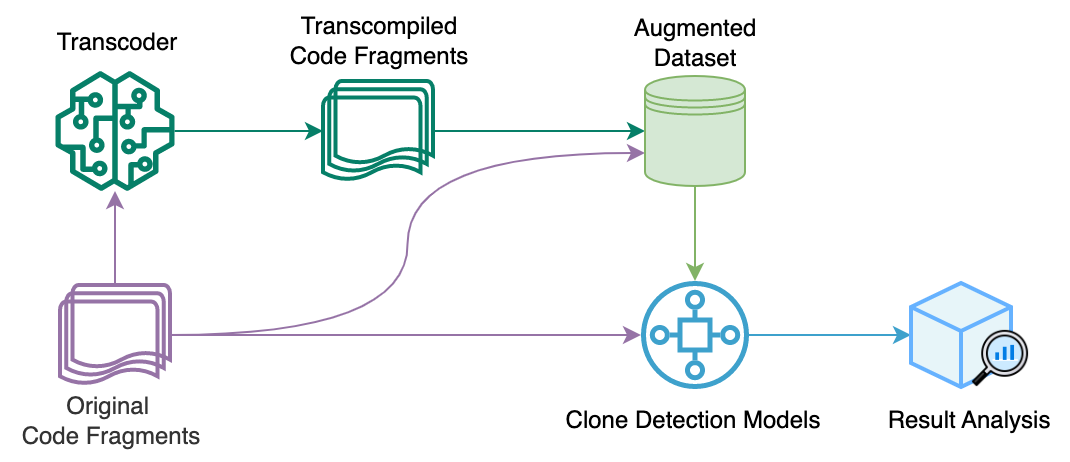}
\caption{Overview of our approach to data augmentation and model training.}
\label{fig:overview}
\end{figure}

\jyoti{
Figure \ref{fig:overview} illustrates the overall approach of our study while examining the impact of data augmentation on cross-language clone detection models. 
First, we convert each code fragment from the original dataset to get another version of it in another language (e.g., Java to Python and vice versa).  Second, the original code fragments and the transcompiled 
fragments are stored in a database to be used 
for the assessment of the impact of augmentation. Third,  the models are trained in two different ways, one that only uses the original dataset and  the other using the augmented dataset. Here we compute the model accuracies and compare them to compute the impact of data augmentation. 

\subsection{Data Augmentation using Transcoder}
In our study, we are particularly interested in generating data for cross-language clone detection models. We used source-to-source translation for this task. To create more data, we leveraged the pre-trained model, Transcoder, to convert codes from one language to another. In our dataset, we had Java and Python. Consequently, we converted all Java codes to Python and Python codes to Java. Since the Transcoder is a pre-trained deep learning model, it is very effective and efficient in generating fragments that share the same semantics as the original ones. Moreover, the pattern learned by the pre-trained models is not based on hand-crafted rules. As a result, the kind of codes it generates are diverse in nature. The details about semantic, syntactic, and computational correctness of the generated codes can be found in
\cite{roziere2020unsupervised}.

\subsection{Mutation Based Augmentation} 
We created another dataset by transforming code fragments using mutation operations to see how the extent of its impact compares with the ones obtained using Transcoder based augmentation. 
\pinku{The operation includes insertion, deletion, swap, duplication of statements, comments, and change in operators\cite{roy2009mutation, svajlenko2019mutation}.
We chose these transformations based on how different types of clones are defined and categorized. 
For example, comments and white spaces are the only differences for Type I clones. Similarly, insertion, deletion, and modification of statements fall into Type III clones. We chose the operations with random order and frequency. These changes were made at random locations in the code fragment\cite{svajlenko2019mutation}.} 
}

\begin{figure}[pt]
\centering
\includegraphics[width=.45\textwidth]{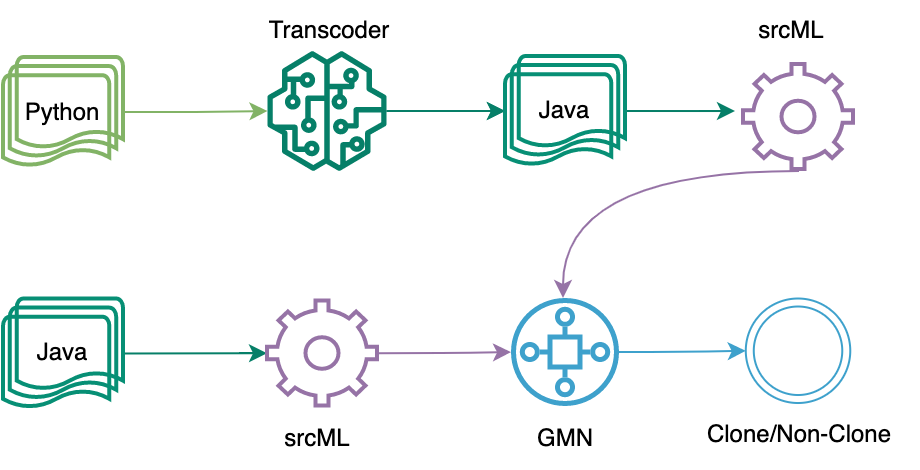}
\caption{Graph matching network combined with a transcompiler.}
\label{fig:GMN_trans}
\end{figure}

\subsection{GMN with Transcompiler}\jyoti{
To demonstrate how transcompiler could be used to extend single-language clone detection tools for cross-language settings, we leverage graph matching network (GMN). GMN is a single-language clone detection model that achieved superior single-language clone-detection performances for all four types of clones\cite{wang2020detecting}. Furthermore, it exploits both the structure and semantics of code fragments which makes it more robust for source code analysis. 

 Figure \ref{fig:GMN_trans} depicts the steps of using GMN with Transcoder. GMN  can only match homogeneous structures, and the embedding space is based on the textual information of tokens for each node. As a result, different structures and texts would generate very different embedding in GMN. We chose Java as the target language for this reason. First, we pass the Python code fragment from a pair through Transcoder to get the converted code representation in Java. Then both of the fragments are passed through srcML to create the AST representation. We use this representation to train the model. We can test it against any fragment following the same conversion steps once the model is trained. 

\begin{figure*}[t!]
\centering
\includegraphics[width=.74\textwidth]{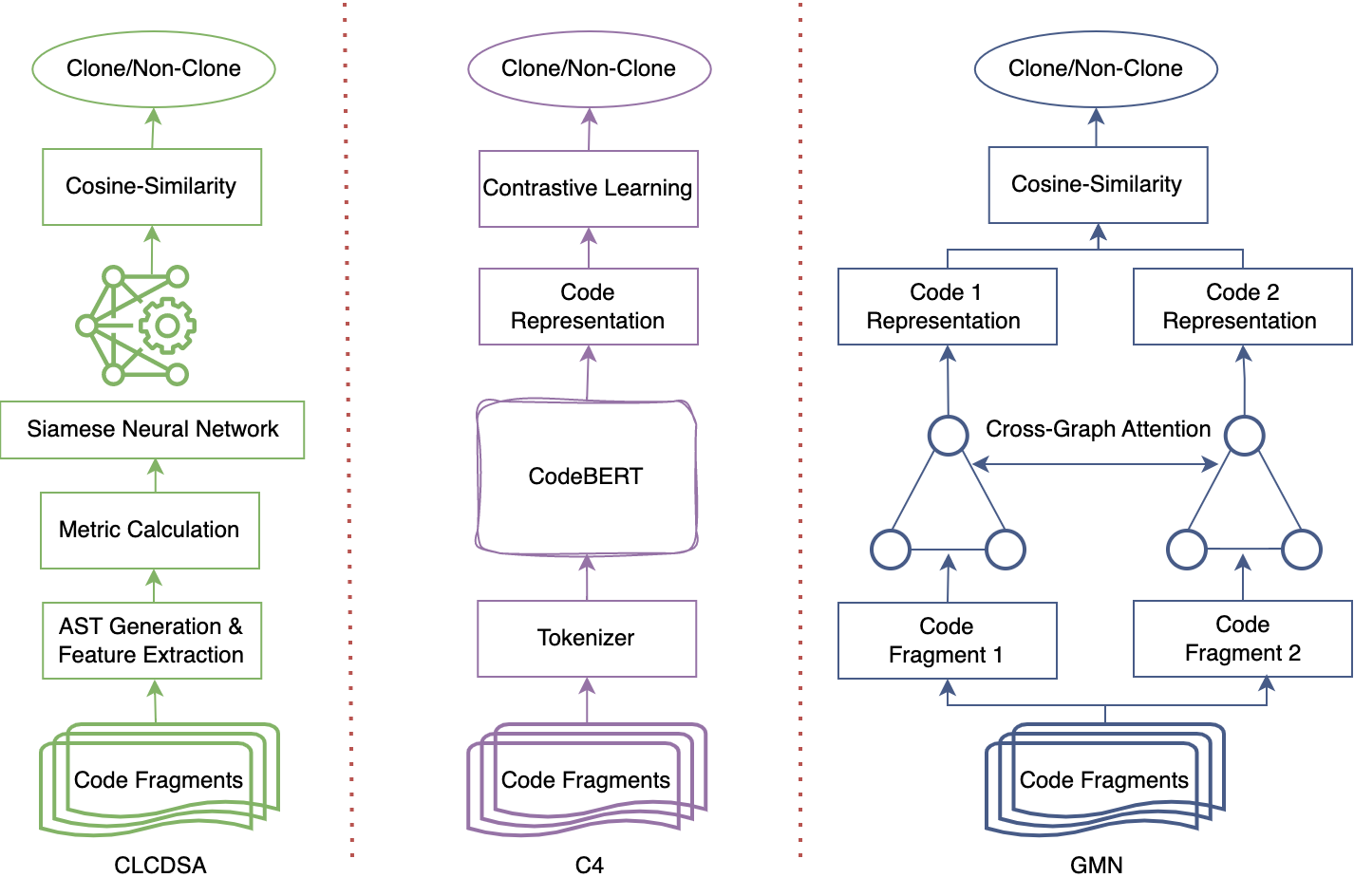}
\caption{Deep learning models in our study. CLCDSA and C4 are existing models for cross-language clone detection. GMN is combined using a transcompiler.
}
\label{fig:models}
\end{figure*}

\begin{table}
\caption{Summaries of Original Dataset}
\label{summary_od}
\centering
\begin{tabular}{ l c c c  c }
\hline
&\multicolumn{2}{c}{ AtCoder(AtC)} &\multicolumn{2}{c}{ GoogleCodeJam(GCJ)}\\
  Metric & Java & Python & Java & Python\\
\hline
\#Problems &1095 &1028 &261 &223\\
\#Average Lines  &55 &19 &73 &57\\
\#Parsable Fragments &14838 &14703 &4341 &1121 \\
\#Unparsable Fragments &620 &11 &5 &2471 \\
\hline

\end{tabular} 
 \end{table}

 \begin{table}[pt]
\caption{Summaries of Transcompiled Dataset}
\label{table:transdata}
\centering
\begin{tabular}{ l c c c  c }
\hline
 &\multicolumn{2}{c}{ AtCoder(AtC)} &\multicolumn{2}{c}{ GoogleCodeJam(GCJ)}\\
  Metric & Java & Python & Java & Python\\
\hline

\#Problems &1027 &1086   &221 &258\\
\#Average Lines  &14 &32 &25 &45\\
\#Parsable Fragments &10024 &5574 &1989 &1550 \\
\#Unparsable Fragments &4684 &2620 &1578 &2620 \\
\hline

\end{tabular} 
 \end{table}
\subsection{Models and Datasets}  
\label{sec:datasets}
The cross-language clone detection models that we used  in our experiments are CLCDSA and C4, and the single-language model that we chose to extend to the cross-language setting is GMN. Figure \ref{fig:models} shows the structures of these  models at a high level.  
We used the popular dataset from the paper that introduced the cross-language clone detection model called CLCDSA\cite{nafi2019clcdsa}. We will refer to this dataset as CLCDSA dataset for convenience. The dataset consists of code fragments from three programming contests that include  AtCoder\footnote{https://atcoder.jp/}, Google CodeJam\footnote{https://codingcompetitions.withgoogle.com/codejam}, and CoderByte\footnote{https://coderbyte.com/}. AtCoder is a programming contest website that originated in Japan, and CodeJam is Google's programming competition. The data for CoderByte is not available in their replication package. 

We chose code fragments of Java and Python programming languages for this study. The study is challenging because these two languages follow different paradigms, such as Java is a statically-typed language, and Python is a dynamically-typed language. 
There are about thirty-eight thousand Java and Python code fragments in the original dataset. We used Transcoder to convert code fragments from the original dataset to get another transcompiled version of it. We create the augmented dataset by combining this data with the original dataset. The summary statistics for original data are given in Table \ref{summary_od}, and the  information regarding transcompiled data is shown in Table \ref{table:transdata}. 
Both of the datasets have many unparsable code fragments.

We followed the pair creation procedure of CLCDSA\cite{nafi2019clcdsa}. Code fragments from the same problem are identified as clones, and code fragments from different problems are non-clones. We had 312,581 clone and non-clone pairs in the original dataset and 1,130,998 in the augmented dataset.

\pinku{We created a smaller version of the dataset by filtering out fragments of less than six lines which is often considered as the minimum granularity for functional clone\cite{roy2008nicad}. Then we randomly selected 60\% of these code fragments to create clone pairs, which we will refer to as the `sample dataset'. This sample dataset is around one-third of the original dataset and consists of 134,623 pairs. All code fragments from the sample dataset were converted using Transcoder. These converted fragments were combined with sample dataset to create the augmented sample dataset. The augmented sample dataset consists a total of 825,050 pairs.}

\pinku{To augment the dataset, clone/non-clone pairs were created after combining the original code fragments with the new TransCoder generated code fragments.  We selected 20\% random pairs from these new pairs and combined them with the pairs from the original dataset. As a result, this augmentation process increased the datasets by 20\% for both the original dataset and sample dataset.}

\pinku{We split the data into 8:1:1 ratio for train, validation, and test set\cite{nafi2019clcdsa,tao2022c4, wang2020detecting}. Since binary classification can be biased with an imbalanced number of items for each class, we maintained a 1:1 ratio for clone and non-clone pairs across all models. We followed the standard procedure to select an equal number of clone and non-clone pairs randomly\cite{tao2022c4}. The models in this study have also used the same ratio and have been trained through this procedure to ensure data balance and reduce bias\cite{nafi2019clcdsa, tao2022c4, wang2020detecting}.} We followed the same procedure while creating the other augmented dataset with the mutation-based code transformations. 
}
\subsection{Evaluation Metrics}
\pinku{Precision, recall, and F1 score  are the most widely used\cite{tao2022c4, wang2020detecting} 
 metrics for clone detection tasks, which are defined as follows.}

\begin{equation}
    Precision = \dfrac{T_P}{T_P+F_P}
\end{equation}
\begin{equation}
    Recall = \dfrac{T_P}{T_P+F_N}    
\end{equation}
\begin{equation}
    F1 { score} = \dfrac{2 \times  Precision \times  Recall}{Precision + Recall}
\end{equation}

Here, $T_P$ is the number of clones classified correctly, and $T_N$ is the non-clones classified correctly. $F_P$ stands for non-clones that were mistaken as clones, and $F_N$ is clones that were classified as non-clones by the model. 
\pinku{Since Transcompiler merely provides us with a code representation, the values of these metrics depend on the clone detection models.} We use F1 score as the ultimate measure for model evaluation.

\subsection{Experimental Settings}
We used the Transcoder model\footnote{https://github.com/facebookresearch/CodeGen} specifically trained with deobfuscation objectives for Java and Python. We used the default settings for the Transcoder as stated in the original paper\cite{roziere2020unsupervised}. We followed the settings in CLCDSA\cite{nafi2019clcdsa} for filtering out clones and non-clones to prepare the dataset. We followed the settings mentioned in the respective papers to train the models\cite{nafi2019clcdsa},\cite{tao2022c4},\cite{wang2020detecting}. 

In the case of C4, we choose a batch size of 4. 
We replicated the model with the original dataset provided in the replication package and this setting and found less than 1\% disagreement between the results.  
\pinku{
We trained and tested the models following existing literature\cite{tao2022c4, wang2020detecting} on a machine using RTX-3080ti.} 
Our codes and datasets are  available   
\href{https://github.com/subrotonpi/clone_transcompiler}{here}.

\section{Results}

\subsection{Answering RQ1:  Impact of data augmentation through source-to-source translation }
To answer the first research question, we divided the clone detection techniques into two groups. The first three techniques in Table \ref{table:results_od} are the deep learning techniques (DL models) used in this study. The last row shows the non-machine learning (non-ML) approaches. \pinku{Among the deep learning techniques, 
CLCDSA has been considered as a reasonable standard baseline, and C4 is the state-of-the-art cross-language clone detection model\cite{mathew2021cross},\cite{tao2022c4}.} The GMN model is the  adapted single-language clone detection model for cross-language clone detection. 

\begin{table}[pt]
\caption{Results on Original Dataset.  The first row shows the deep learning models and the second row shows Non-ML Models}
\label{table:results_od}
\centering
\begin{tabular}{l l c c c c} 
 \hline
 &Model & Precision & Recall & F1 \\
 \hline
\multirow{3}{5em}{DL Models} 
&CLCDSA &0.49 &\textbf{0.99} &0.66\\
&C4 &\textbf{0.95} &0.96 &\textbf{0.96}\\
&GMN &0.90 &0.92  &0.91\\
\hline
\multirow{2}{5em}{Non-ML Models}  
&CLCMiner &0.36 &0.57 &0.44\\
&COSAL &0.55 &0.89 &0.68\\
\hline
\end{tabular}
\end{table}

\begin{table}[pt]
\caption{Results on Augmented Dataset. Delta shows the difference with original dataset in F1 score from (Table \ref{table:results_od})}
\label{table:results_ad}
\centering
\begin{tabular}{l c c c c c} 
 \hline
 Model & Precision & Recall & F1 &$\Delta$ \\
 \hline

CLCDSA &0.53 &0.98 &0.69 &\textbf{3\%}\\
C4 &\textbf{0.97} &\textbf{0.99} &\textbf{0.98} &2\%\\
GMN &0.93 &0.95 &0.94 &\textbf{3\%}\\

\hline
\end{tabular} 
\end{table}  

We trained each model with original and augmented datasets. We also trained with the smaller version of each dataset to understand the degree of impact on the models over dataset size. As a result, this combination makes four sets of datasets in total. Table \ref{table:results_od} shows the precision, recall, and F1 score for each of the DL models we trained with the original dataset. Among the models, C4 has the best precision and F1 score, and CLCDSA has the highest recall score. Table \ref{table:results_ad} shows the scores for these models when trained with the augmented dataset. All of the models show an improvement of 2\% to 3\% in terms of F1 score. CLCDSA and GMN showed a higher increase compared to C4 in the F1 score.

Table \ref{table:results_ds} shows that for the smaller sample dataset (as described in Section \ref{sec:datasets}). Model C4 again has the highest F1 score and CLCDSA still maintains its high recall value. Table \ref{table:results_ad_small} shows the result on the augmented sample dataset. In this case, CLCDSA and GMN show a 2\% increase in the F1 score.  
The overall F1 scores for all models and the performance boosts appear to be small when compared to the results on the original dataset, which is expected as the performances of deep learning models often suffer from the lack of data. However, in all cases, the results show that the Transcoder-based data augmentation can effectively improve the clone detection performances of the deep learning  models.
\begin{tcolorbox}[left=0pt,right=0pt,top=0pt,bottom=0pt,boxrule=0.1pt,colback=black!5!white,colframe=black!75!black]
In summary, our experimental results show that the transcompiler-based data augmentation can substantially increase deep learning models' performance for cross-language clone detection (e.g., for some models the F1 score improves about 3\% on benchmark dataset). 
\end{tcolorbox} 


\begin{table}[pt]
\caption{Results on Sample Dataset}
\label{table:results_ds}
\centering
\begin{tabular}{l c c c c} 
 \hline
Model & Precision & Recall & F1\\
 \hline

CLCDSA &0.50 &\textbf{0.99} &0.66\\
C4 &\textbf{0.92} &0.95 &\textbf{0.93}\\

GMN &0.89 &0.91 &0.90\\
\hline
\end{tabular} 
\end{table}  
\begin{table}[pt]
\caption{Results on Augmented Sample Dataset. Delta is the difference   in F1 score from (Table  \ref{table:results_ds})}
\label{table:results_ad_small}
\centering
\begin{tabular}{l c c c c} 
 \hline
Model & Precision & Recall & F1 &$\Delta$\\
 \hline

CLCDSA &0.54 &0.93 &0.68 &\textbf{2\%}\\
C4  &\textbf{0.93} &\textbf{0.96} &\textbf{0.94} &1\%\\
GMN &0.91 &0.93 &0.92 &\textbf{2\%}\\
\hline
\end{tabular} 
\end{table}

\subsection{Answering RQ2: Comparing mutation based augmentation with transcompiler based augmentation} 
We found that augmentation with source-to-source translation worked well for both large and small datasets. However, at this point, a natural question is whether we could find the same level of performance boost using simple mutation based data augmentation. To answer this research question, we examined the performances\cite{svajlenko2019mutation} of the models on the 
mutation based augmented dataset. We followed the methodology described earlier (Section \ref{sec:datasets}) to create the mutation based  augmented dataset, and re-trained all of the models.  We also created a smaller sample of this dataset and maintained the same number of pairs and the ratio between clone and non-clone pairs as we did for RQ1.

\begin{table}[htbp]
\caption{Results on mutation based dataset. Delta is the change in F1 score  compared with transcompiler approach (Table \ref{table:results_ad}, \ref{table:results_ad_small})}
\label{table:results_non_sem}
\centering
\begin{tabular}{l l c c c c} 
 \hline
Dataset Size &Model & Precision & Recall & F1 &$\Delta$\\
 \hline

\multirow{3}{5em}{Original} 
&CLCDSA &0.52 &0.97 &0.68 &-1\%\\
&C4  &\textbf{0.97} &\textbf{0.98} &\textbf{0.97} &-1\% \\
&GMN &0.89 &0.90 &0.91 &\textbf{-3\%}\\
\hline
\multirow{3}{5em}{Small sample} 
&CLCDSA &0.50 &\textbf{0.98} &0.67  &-1\%\\
&C4  &\textbf{0.930} &0.944 &\textbf{0.936} &-0.4\% \\
&GMN  &0.88 &0.90 &0.89 &\textbf{-3\%}\\
\hline
\end{tabular} 
\end{table}  


The rightmost column shows the change in the F1 score for the mutated dataset compared to our transcompiler based approach. It is evident from Table \label{table:results_non_sem} that for all  models, performance decreased from 1\% to 3\% in the case of a larger dataset, and it dropped 3\% for GMN for the smaller sample dataset. C4 managed to perform almost similar for both the original dataset and the smaller sample dataset. In the case of the original dataset, it shows only a decline of 1\% while maintaining high accuracies. We believe that the power of C4 lies in the CodeBERT model, as it is primarily dependent on the token sequences instead of the structure or semantics of code fragments.

We further investigate the reason why mutation based augmentation fails to provide a performance boost whereas the transcompiler based approach appears to work. 

\subsection*{Why does the Transcompiler Based Approach Work?}

To this end, we believe the quality of data impacts a model's performance to a large extent. 
Although mutation based modification of source codes does not hamper the token quality, 
it may create unparsable code, and  alternate the syntax and sequence of execution. Consequently, it affects the AST tree structure. 
Therefore, it may initially appear that the new code fragments that are  created by the  mutation based augmentation are more diverse than the ones created by the transcompiler based approach. To examine this we computed the ASTs of all the code fragments and computed a  similarity score between ASTs obtained from the original code   and augmented code.  
 Specifically, we compared the metric average root-to-leaf distance for the ASTs (i.e., the mean of the distances from the root to all leaf nodes for each ASTs), as shown in Table \ref{table:root_stat}.

\begin{table}[pt]
\caption{AST distance statistics.}
\label{table:root_stat}
\centering
\begin{tabular}{l l l c c c c} 
 \hline
Dataset Size & &Original & Transcompiled &  Mutated\\
 \hline
\multirow{3}{5em}{Large} 
&Mean &17.93 &16.53 &17.85 \\
&SD  &0.244 &0.384 &0.277\\
\hline
\multirow{3}{5em}{Small} 
&Mean &17.94 &16.55 &17.86 \\
&SD  &0.249 &0.407 &0.281\\
 \hline
\end{tabular} 
\end{table} 

  Contrary to the initial assumption, we now can see that  the variation appears to be more in the transcompiler based augmented data compared to the mutation based augmented dataset.  We investigated the reason for this and found the code generated from Python to Java by Transcoder is shorter compared to the mutated fragments (Java to Java). 
The AST lengths are quite similar to the original code in case of random perturbation unless many statements were added and deleted. Figure \ref{example_trans_rand} shows examples of a code fragment that can read a text file. When it is transcompiled from Java to Python using Transcoder, the number of lines is reduced. As evident from the last fragment in the example, the random application of the mutation operation modifies the code by deleting and swapping some lines of code. Lines 6 and 7 from  (a) were swapped and lines 3 and 9 were deleted by the operations. This observation shows that the data produced by transcompilers are more diverse in nature. These code fragments have inherent variation in their structure since the pre-trained model was trained on a large corpus of real-world data. This tends to capture more variation from the data and use that in inference. This explains the success of transcompiler based approach to some extent as removing redundancies and increasing the diversity in datasets tends to improve the performance of machine learning models\cite{gong2019diversity}.

\begin{table}[pt]
\caption{Mean Absolute Difference between the Average Root-to-Leaf Distances for  the ASTs corresponding to the Original and Newly Created Code Fragments }
\label{table:root_diff}
\centering
\begin{tabular}{l c c} 
 \hline
 &Original vs. Transcompiled &Original vs. Mutated \\
\hline

Mean  &1.3 &0.09 \\
SD &0.37 &0.17\\
 \hline
\end{tabular} 
\end{table} 

\begin{tcolorbox}[left=0pt,right=0pt,top=0pt,bottom=0pt,boxrule=0.1pt,colback=black!5!white,colframe=black!75!black]
In summary,  the mutation-based data augmentation may not boost the model performances and thus careful data augmentation such as Transcoder based ones are important. A potential reason for the success of  Transcoder based data augmentation is their capability of increasing data diversity while preserving the code semantics. 
\end{tcolorbox}


 \subsection{Answering RQ3: Extending single-language model for cross-language clone detection}
Here we examined GMN in more detail. Recall that although GMN is known to provide high accuracy for single-language settings, it   
has not previously been used for cross-language clone detection. 
Our experiment shows GMN outperformed the baseline model CLCDSA by a high margin. It showed a very robust result,  for all the cases examined in Tables \ref{table:results_od}--\ref{table:results_ad}. 
Since GMN relies on the code  structure more than the other two models, it showed a larger decline in performance when used with the mutation based augmentation method. The original implementation of GMN did not consider unparsable codes. We exploited srcML\footnote{https://www.srcml.org/about.html} parser to create a graph for all types of codes generated by transcompiler irrespective of whether they are   parsable or not. 

 \begin{table}
\caption{Results on only parsable codes vs. all. 
}
\label{table:results_parsable_all}
\centering
\begin{tabular}{c c c c} 
 \hline
 Parser &Metrics  &Original Dataset &Augmented Dataset\\
 \hline
  \multirow{3}{4em}{Only Parsable} 
  &Precision &0.90 &0.92\\
  &Recall &0.92 &0.94\\
  &F1  &0.91 &0.93\\ 
 \hline
\multirow{3}{4em}{All} 
&Precision &0.90 &0.93\\
  &Recall &0.92 &0.95\\
  &F1  &0.91 &\textbf{0.94}\\ 

\hline
\end{tabular} 
 \end{table}
 
 Table \ref{table:results_parsable_all} shows the results  on original and augmented datasets. The top rows show the result obtained using only the code fragments that were parsable and the bottom rows show the result for all code fragments (i.e., including unparsable ones). We can see for augmentation, taking all types of code fragments helped GMN to learn better representations. 
 
GMN takes significantly less time than C4. The reason is the number of parameters and fine-tuning time required by CodeBERT\cite{feng2020codebert}. The number of trainable parameters for GMN is 123,101, whereas C4 has 172,503,552, which is 1401 times higher. To train the models, GMN takes around one-fourth time of C4. Hence we believe that GMN can be an excellent choice in a resource-constraint environment and it now appears as a competitive candidate  to be examined further for cross-language clone detection in future research.

\begin{tcolorbox}[left=0pt,right=0pt,top=0pt,bottom=0pt,boxrule=0.1pt,colback=black!5!white,colframe=black!75!black]
  In summary, our results show that single-language clone detection models may be used in a combination with transcompilers to create a pipeline for effective cross-language clone detection. However, such an adaptation may require additional steps depending on the specifications of the single-language clone detection models being extended.  
\end{tcolorbox}
 
\section{Discussion}

\subsection{Generalizability of Our Approach} 
The quality of the data augmentation using source-to-source translation depends on the choice of transcompiler. This study considered only Java and Python programming languages as a use case. These two languages are way different from one another compared to the languages such as C\# and C++ which have more  lexical features in common. The performance of  the Transcoder depends on the agreement of the dynamic or static nature of the taken languages. The more common characteristics two different language shares, the better outcome is possible from the Transcoder. \pinku{However, TransCoder is a pre-trained model and it can convert a code fragment with a fraction of a second and therefore, can be used for large-scale clone detection tasks.}

We showed that our approach is efficient as the pre-trained model does not require any kind of fine-tuning and can be used for inference directly for code conversion. Moreover, the amount of time required for model inference is negligible. These characteristics make this approach ideal and generalized for cross-language data augmentations. \pinku{Our experimental results show consistent improvements for all models even when they already achieved over 90\% F1 score. This indicates that practitioners may want to use such data augmentation techniques to their fullest potential.} Additionally, the experimental result also shows success in the adaption of a single-language clone detection model through the aid of a transcompiler. Although we chose a specific graph neural network for this task, 
the concept generalizes to other deep learning based single-language clone detection model. 

\setbox0=\hbox{%
\begin{minipage}{\linewidth}
\begin{lstlisting}[columns=flexible,language=Java,numbers=none, style=code_table]
public static readTextFile( ) {
    try {
      File f = new File("textfilename.txt");
      Scanner sc = new Scanner(f);
      while(sc.hasNextLine( )) {
        String st = sc.nextLine( );
        System.out.println(st);
      }
      sc.close( );
    } catch(FileNotFoundException e) {
      e.printStackTrace( );
    }
  }

\end{lstlisting}
\end{minipage}
}
\savestack{\listingA}{\box0}

\setbox0=\hbox{%
\begin{minipage}{\linewidth}
\begin{lstlisting}[columns=flexible,language=Python,numbers=none,style=code_table]
def read_textfile( ) :
    with open('textfilename.txt', 'r') as f :
        with f :
            for st in f :
                print(st)
\end{lstlisting}
\end{minipage}
}
\savestack{\listingB}{\box0}
\setbox0=\hbox{%
\begin{minipage}{\linewidth}
\begin{lstlisting}[columns=flexible,language=Java,numbers=none,style=code_table]
public static readTextFile( ) {
    try {
      Scanner sc = new Scanner( );
      while(sc.hasNextLine( )) {
        System.out.println(st);
        String st = sc.nextLine( );
      }
    } catch(FileNotFoundException e) {
        e.printStackTrace( );
    }
  }
\end{lstlisting}
\end{minipage}
}
\savestack{\listingC}{\box0}
\begin{figure}[t!]
\stackinset{l}{-5pt}{t}{\llength}{}{\listingA} 
\textbf{\footnotesize (a) Java code for reading a text file} 

\stackinset{l}{-5pt}{t}{\llength}{}{\listingB} 
\textbf{\footnotesize (b) Python code by Transcoder}

\stackinset{l}{-5pt}{t}{\llength}{}{\listingC} 
\textbf{\footnotesize (c) Java code by  mutation} 

\caption{Example of codes modified differently. The code in (b) is transcompiled through Transcoder from (a), and code in (c) is generated by applying random mutation operations.}
\label{example_trans_rand}
\end{figure}

\subsection{Threats to validity}
Our method relies on a pre-trained Transcoder model. We used it to convert code fragments from one language to another language. This may introduce the inherent bias in our study. Transcoder is trained on a large corpus of codes. It outputs varied levels of computationally correct codes depending on the choice of programming language. 
 In our technique, the languages that could be supported are limited by the ones supported by the transcompilers. Nevertheless, the rapid development of pre-trained models and different fine-tuning approaches for transcompilers may overcome these limitations.


We used the dataset provided by CLCDSA paper\cite{nafi2019clcdsa}, which is collected from programming competitions. As a consequence, they  can be very different from real-world software systems. Moreover, many of these problems are complex, and the  solutions require a well-thought approach. Therefore, the code fragments are likely to be diverse both syntactically and semantically. Hence it would be interesting to explore our proposed method in real-life software systems.

We considered only two languages for our study whereas the original dataset has code fragments from more languages. We observed that other studies considered different sets of programming languages from this dataset. For example, CLCDSA\cite{nafi2019clcdsa} used three languages in their study, C4 used four languages\cite{tao2022c4}, and COSAL\cite{mathew2021cross} used only Java and Python. 
Examining how the performances vary across different languages could be an interesting avenue for future research.

\section{Conclusion}
Clone detection is a widely studied field in software engineering research. However, techniques for cross-language clone detection are not extensively explored compared to their single-language counterpart. 
Here we proposed a novel data augmentation technique that uses a transcompiler (a pre-trained deep learning model for source-to-source translation) for cross-language clone detection. Our experiment shows that such data augmentation improves the performances of cutting-edge cross-language clone detection models. We also exploited a single-language model for detecting cross-language clones through source-to-source translation. The performance of the single-language model surpassed the current baseline by a large margin. This opens up an opportunity to further examine  single-language clone detection models to understand how the transcompilers could be combined more effectively to achieve even better cross-language clone detection performances.  In the future, we envision applying explainable AI techniques to further explain the power of data augmentation and to develop techniques to select the augmented data  appropriately to further enhance the effectiveness of our approach.

 \section{Acknowledgement}
 This work was supported by   NSERC Discovery, CFI-JELF, and NSERC CREATE graduate program on Software Analytics Research (SOAR) grants.

\bibliographystyle{ieeetr}
\bibliography{reference.bib}
\end{document}